\documentclass[12pt,a4paper]{article}

\usepackage[T1]{fontenc}
\usepackage[utf8]{inputenc}
\usepackage{amsmath,amssymb,amsfonts}
\usepackage{graphicx}
\usepackage{bm}
\usepackage{booktabs}
\usepackage{array}
\usepackage{hyperref}
\usepackage{geometry}
\usepackage{microtype}
\usepackage{placeins}
\usepackage{xcolor}
\usepackage{titlesec}
\usepackage{caption}
\usepackage{float}

\hypersetup{
  hidelinks,
  pdftitle={Light-front transverse profiles of scalar and spin-two chromoelectric EMT responses in near-threshold charmonium probes},
  pdfauthor={Arkadiy I. Syamtomov},
  pdfsubject={Light-front transverse chromoelectric EMT profiles in near-threshold charmonium probes},
  pdfkeywords={energy-momentum tensor, trace anomaly, charmonium, light-front distributions, gravitational form factors}
}
\titleformat{\section}
  {\normalfont\normalsize\bfseries}
  {\thesection}{0.7em}{}

\newcommand{\dd}{\mathrm d}
\newcommand{\GeV}{\mathrm{GeV}}
\newcommand{\fm}{\mathrm{fm}}

\newcommand{\calA}{\mathcal A}

\newcommand{\Dperp}{\bm\Delta_\perp}

\begin{document}

\title{Light-front transverse profiles of scalar and spin-two chromoelectric EMT responses\\ in near-threshold charmonium probes}

\author{Arkadiy I.~Syamtomov\thanks{Corresponding author: \href{mailto:arkady.syamtomov@gmail.com}{\texttt{arkady.syamtomov@gmail.com}}}}
\date{\small\textit{Bogolyubov Institute for Theoretical Physics,\\
National Academy of Sciences of Ukraine,\\
Kyiv, Ukraine}}
\maketitle

\begin{abstract}
I construct light-front transverse profile functions for the scalar and spin-two chromoelectric energy-momentum-tensor responses selected by compact-charmonium probes of the proton.  The scalar branch is the anomaly and sigma amplitude, while the spin-two branch contains the gravitational combination $3B_i(t)-D_i(t)=6J_i(t)-3A_i(t)-D_i(t)$ rather than $A_i(t)$ alone.  In the Drell--Yan frame, these two amplitudes define normalized transverse response profiles whose integrated ratio reproduces the forward light-front spin-two/scalar strength, while the finite transverse-distance behavior is sensitive to the relative scalar and gravitational slopes.  The construction gives a spatial representation of the chromoelectric EMT projection without interpreting a measured near-threshold slope as a model-independent static three-dimensional mass radius.
\end{abstract}


\section{Introduction}

Near-threshold charmonium production is often discussed as a probe of gluonic gravitational structure.  The short-distance reason is well known: for a compact heavy quarkonium state, the leading soft interaction is governed by the chromoelectric operator of the QCD multipole expansion~\cite{Peskin1979,BhanotPeskin1979}.  What is less direct is the spatial interpretation of the measured $t$ dependence.  A slope in near-threshold production is not automatically a static three-dimensional mass radius; in a relativistic light-front description the natural spatial variable is the transverse impact parameter in the Drell--Yan frame, $\Delta^+=0$ and $t=-\Dperp^2$~\cite{Burkardt2000,Diehl2003,FreeseMiller2021,FreeseMiller2022,BurkertRMP2023}.

The operator matching and off-forward spin-two contraction are taken from the operator analysis of Ref.~\cite{Syamtomov2026EMT}, which determined the chromoelectric scalar and spin-two content.  I do not repeat that derivation here.  The result needed in the present paper is that the compact-charmonium probe separates into a dominant scalar anomaly and sigma amplitude and a subleading traceless spin-two amplitude.  Forward on the light front, the spin-two term is controlled by partonic plus-momentum fractions.  Off forward, the chromoelectric spin-two contraction is not an $A_i(t)$ form-factor slope alone; it contains the combination $3B_i(t)-D_i(t)$, equivalently $6J_i(t)-3A_i(t)-D_i(t)$.

This paper asks a complementary question.  Once the scalar and spin-two amplitudes are known, how should their transverse profile shapes be represented?  The answer is a shape-strength separation.  The forward normalizations are fixed by the chromoelectric EMT matching, while the transverse profiles are obtained from normalized off-forward amplitude shapes.  This preserves the light-front density interpretation where it is applicable and avoids assigning an independent density meaning to scheme-dependent bookkeeping terms such as $\bar C_i(t)$.

The aim is therefore not to introduce light-front gravitational densities as new objects.  Those distributions and their limits are part of an established EMT/GFF literature, from impact-parameter and light-front density constructions to mechanical EMT distributions and modern reviews~\cite{Burkardt2000,Diehl2003,FreeseMiller2021,FreeseMiller2022,BurkertRMP2023,PolyakovWeiss1999,PolyakovSchweitzer2018,LorceMoutardeTrawinski2019}.  The new element is to apply that language to the corrected chromoelectric EMT projection and to separate three quantities that are easily conflated: the forward scalar/spin-two strength, the off-forward form-factor shape, and the transverse localization scale.

\section{Chromoelectric amplitudes}
\label{sec:amplitudes}

The operator matching of Ref.~\cite{Syamtomov2026EMT} gives the scalar amplitude in the anomaly/sigma basis
\begin{equation}
  N_0(t)=C_{\calA}\,\calA_N(t)+C_\sigma\,\sigma_N(t),
\label{eq:N0}
\end{equation}
where
\begin{equation}
  \calA=\frac{\beta(g)}{2g}[F^2]_R+\gamma_m\sum_q m_q\bar q q,
  \qquad
  \sigma_N(t)=\frac{1}{2M}\langle P'|\sum_q m_q\bar q q|P\rangle .
\label{eq:scalar_basis}
\end{equation}
At $t=0$, $\calA_N(0)=M-\Sigma_N$.  The scalar branch is therefore not a pure beta-function trace at finite quark mass.  In the numerical profiles below I use a common normalized scalar shape for $\calA_N(t)$ and $\sigma_N(t)$, as in the benchmark scan of Ref.~\cite{Syamtomov2026EMT}.  This keeps the present paper focused on the transverse profile construction rather than on a separate scalar-dispersion analysis.

For the spin-two branch, the relevant helicity-conserving contraction in the symmetric $\xi=0$ frame is
\begin{equation}
  N_2(t)=\sum_{i=q,g} C_i^{(2)}
  \left[
  \frac{3}{4}M A_i(t)+\frac{t}{16M}\left(3B_i(t)-D_i(t)\right)
  \right].
\label{eq:N2_ABD}
\end{equation}
Using the standard EMT relation
\[
J_i(t)=\frac12\left[A_i(t)+B_i(t)\right],
\]
which is the off-forward form-factor version of Ji's angular-momentum
sum rule~\cite{JiSpin}, the same contraction may be written as
\[
3B_i(t)-D_i(t)=6J_i(t)-3A_i(t)-D_i(t).
\]
This form is useful because it displays explicitly that the
off-forward spin-two chromoelectric response depends on the
angular-momentum form factor \(J_i(t)\) and the mechanical
form factor \(D_i(t)\), and not on the momentum form factor
\(A_i(t)\) alone. Equations~\eqref{eq:N0}--\eqref{eq:N2_ABD} are the only operator-level input for the present profile construction. In particular, the $\bar C_i(t)$ term cancels from the traceless spin-two contraction and is not used to define a physical chromoelectric transverse response profile.

For the transverse-profile construction below, the absolute forward strengths and the
off-forward shapes play different roles. The former fix the integrated weights of the
scalar and spin-two profiles, while the latter determine their \(b\)-dependence. I therefore
separate the forward spin-two/scalar strength through
\begin{equation}
  R_{2/0}^{\rm LF}(0)=\frac{N_2(0)}{N_0(0)}.
\label{eq:Rforward}
\end{equation}
For the central CT18NNLO input at $Q_{\rm LF}=2~\GeV$ and matching scale $\mu_h=2~\GeV$, Ref.~\cite{Syamtomov2026EMT} gives $R_{2/0}^{\rm LF}(0)=0.147$.  Varying the matching scale over $2$--$10~\GeV$ gives a representative range $R_{2/0}^{\rm LF}(0)\simeq0.10$--$0.15$.  In this paper I use the convention $N_0(0)=1$ and $N_2(0)=R_{2/0}^{\rm LF}(0)$, so that all remaining information resides in normalized shapes.

\section{Light-front transverse response profiles}
\label{sec:profiles}

In the Drell--Yan frame,
\begin{equation}
  \Delta^+=0,\qquad t=-\Dperp^2,
\end{equation}
the transverse momentum transfer is Fourier conjugate to the light-front
impact parameter. This is the standard impact-parameter representation
of zero-skewness GPDs and form factors, developed in Refs.~\cite{Burkardt2000,Diehl2003}
and used in the light-front EMT density construction of
Refs.~\cite{FreeseMiller2021,FreeseMiller2022,BurkertRMP2023}.
Accordingly, a normalized form-factor or response-amplitude shape \(F(t)\)
is represented by the two-dimensional Fourier--Bessel profile
\begin{equation}
  \rho_F(b)=\frac{1}{2\pi(\hbar c)^2}\int_0^\infty \dd Q\,Q\,
  J_0\!\left(\frac{Qb}{\hbar c}\right)F(-Q^2),
  \qquad Q=|\Dperp|.
\label{eq:FB}
\end{equation}
The factor $(\hbar c)^{-2}$ appears because $Q$ is measured in GeV while $b$ is shown in fm.  For $F(0)=1$, the normalization is $\int \dd^2b\,\rho_F(b)=1$.

The scalar and spin-two chromoelectric amplitudes are first normalized as
\begin{equation}
  \widehat N_0(t)=\frac{N_0(t)}{N_0(0)},\qquad
  \widehat N_2(t)=\frac{N_2(t)}{N_2(0)}.
\label{eq:normalized_shapes}
\end{equation}
A transverse Fourier transform also requires a short-distance completion of the low-$|t|$ GFF model.  I use a common smooth resolution factor
\begin{equation}
  W_\Lambda(t)=\frac{1}{(1-t/\Lambda_{\rm res}^2)^2},\qquad
  \Lambda_{\rm res}=4~\GeV,
\label{eq:Wres}
\end{equation}
applied to both scalar and spin-two shapes.  This factor is not an additional
dynamical form factor.  It is a common ultraviolet completion of the profile
transform, introduced only to avoid assigning physical meaning to arbitrarily
short transverse distances where the low-\(|t|\) continuation is not controlled.
Using the same factor for both branches prevents the short-distance completion
from generating an artificial scalar--spin-two difference.  Since
\(W_\Lambda(0)=1\), it does not alter the forward normalizations or the
large-\(B\) limit of the cumulative ratio.  It only avoids assigning physical meaning to arbitrarily short transverse distances where the local low-energy continuation is not controlled.  The numerical value $\Lambda_{\rm res}=4~\GeV$ is not fitted and is used only to make the Fourier transform well behaved at distances below the domain of the low-$|t|$ GFF model.

This leads to the scalar and spin-two response profiles
\begin{equation}
  \rho_0(b)=N_0(0)\,\rho_{\widehat N_0 W_\Lambda}(b),\qquad
  \rho_2(b)=N_2(0)\,\rho_{\widehat N_2 W_\Lambda}(b).
\label{eq:profile_defs}
\end{equation}
Here the prefactors carry the physical forward strengths, whereas the
Fourier transforms describe the normalized transverse shapes.
Equation~\eqref{eq:profile_defs} is a profile representation of the response amplitude selected by the chromoelectric probe.  It should not be read as a universal mass density of the proton.

A useful integrated diagnostic is obtained by accumulating both profiles inside the
same transverse disk of radius \(B\).  The variable \(B\) is therefore an observation
scale in the light-front impact-parameter plane: small \(B\) probes the central
part of the response profile, while increasing \(B\) includes more of the transverse
tail.  This motivates the cumulative ratio
\begin{equation}
  R_{2/0}(B)=
  \frac{\displaystyle\int_0^B 2\pi b\,\dd b\,\rho_2(b)}
       {\displaystyle\int_0^B 2\pi b\,\dd b\,\rho_0(b)}.
\label{eq:cumulative}
\end{equation}
By construction,
\begin{equation}
  \lim_{B\to\infty}R_{2/0}(B)=R_{2/0}^{\rm LF}(0).
\label{eq:cumulative_limit}
\end{equation}
Thus \(R_{2/0}(B)\) measures how the spin-two/scalar strength ratio builds up
as a function of transverse distance.  Its finite-\(B\) behavior depends on the
relative scalar and spin-two slopes, while the \(B\to\infty\) limit is fixed by
the forward matching.

\section{Scalar and spin-two profile shapes}
\label{sec:central}

The numerical model is deliberately minimal.  It uses the forward normalization and central GFF-shape inputs of Ref.~\cite{Syamtomov2026EMT}, informed by recent lattice and dispersive determinations of nucleon gravitational form factors~\cite{HackettPefkouShanahan2024,CaoGuoLiYao2025}, but does not repeat the operator derivation or attempt a fit to photoproduction data.  For the scalar shape I use the representative scalar trace slope
employed in the finite-\(t\) benchmark of Ref.~\cite{Syamtomov2026EMT},
\(r_s=0.97\,{\rm fm}\), implemented here as a normalized dipole.
This parameter fixes the low-\(|t|\) falloff of the scalar profile only;
it is not interpreted as a model-independent static mass radius. The spin-two shape uses dipole forms for $A_g(t)$, $J_g(t)$, and $D_g(t)$ in the central $\mu_h=2~\GeV$ gluonic matching case.  At this matching scale $C_q^{(2)}=0$, so the central spin-two profile is gluon dominated; evolved quark admixtures are part of the forward-scale uncertainty inherited from Ref.~\cite{Syamtomov2026EMT}.  The profile plots use this central gluon-dominated $\mu_h=2~\GeV$ shape; the matching-scale band shown below is applied as a forward-normalization uncertainty and does not attempt to model possible evolved quark-sector shape differences.  A full treatment of evolved quark-sector transverse shapes would require quark and gluon $A_i$, $J_i$, and $D_i$ profiles at the same scale and is beyond the scope of this profile illustration.

The numerical inputs used for this representative profile construction are summarized in Table~\ref{tab:inputs}.

\begin{table}[H]
\centering
\caption{Central numerical inputs for the LF profile benchmark.  The GFF shape parameters are the central gluon-sector entries used in the finite-$t$ benchmark of Ref.~\cite{Syamtomov2026EMT}.}
\label{tab:inputs}
\begin{tabular}{lll}
\toprule
Quantity & Value & Role \\
\midrule
$M$ & $0.9383~\GeV$ & proton mass \\
$\Sigma_N$ & $59~\mathrm{MeV}$ & sigma-term input in scalar denominator \\
$\alpha_s^{(3)}(2~\GeV)$ & $0.3066$ & scalar coefficient input \\
$R_{2/0}^{\rm LF}(0)$ & $0.1470$ & central forward spin-two/scalar strength \\
$r_s$ & $0.97~\fm$ & scalar dipole slope parameter \\
$A_g(0),\Lambda_A$ & $0.4134,\ 1.262~\GeV$ & gluon momentum GFF shape \\
$J_g(0),\Lambda_J$ & $0.2520,\ 1.399~\GeV$ & gluon angular-momentum GFF shape \\
$D_g(0),\Lambda_D$ & $-2.57,\ 0.538~\GeV$ & gluon mechanical GFF shape \\
\bottomrule
\end{tabular}
\end{table}

To isolate the effect of the different slopes without introducing a
multi-parameter fit, I model each GFF by the same dipole form,
\begin{equation}
  F_i(t)=\frac{F_i(0)}{(1-t/\Lambda_i^2)^2}.
\label{eq:dipole}
\end{equation}
The dipoles are used only as smooth low-\(|t|\) shape parametrizations;
their forward values and pole masses are the benchmark inputs listed in
Table~\ref{tab:inputs}.
The scalar profile is represented by the same dipole form with $\Lambda_s^2=12/r_s^2$ after converting $r_s$ to $\GeV^{-1}$.  This notation follows the conventional slope parameter of the scalar form factor; it is not used as a static radius.

\begin{figure}[H]
\centering
\includegraphics[width=0.66\textwidth]{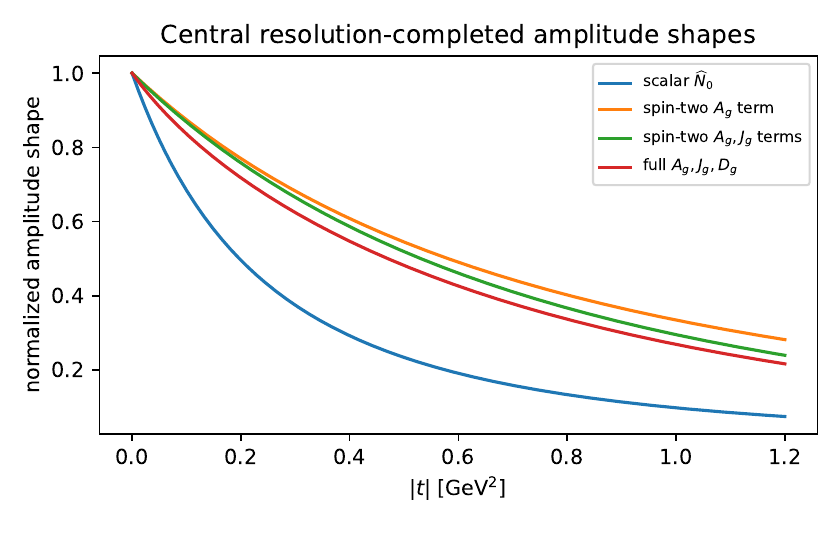}
\caption{Central normalized scalar and spin-two amplitude shapes after the common resolution factor of Eq.~\eqref{eq:Wres}.  The spin-two curves show the $A_g$ term alone, the $A_g,J_g$ truncation, and the full $A_g,J_g,D_g$ contraction.}
\label{fig:shapes}
\end{figure}

Figure~\ref{fig:shapes} shows the normalized resolution-completed $t$-space shapes.  The scalar shape falls faster than the central spin-two shape, while the $D_g$ term changes the finite-$t$ spin-two numerator without affecting its forward normalization.  This is the same operator content as in Ref.~\cite{Syamtomov2026EMT}, now expressed as input for the transverse profile construction.

Selected numerical values for these normalized shapes are listed in Table~\ref{tab:central_numbers}.

\begin{table}[H]
\centering
\caption{Selected normalized resolution-completed scalar and spin-two amplitude shapes for the central benchmark.}
\label{tab:central_numbers}
\begin{tabular}{lccc}
\toprule
Quantity & $|t|=0.2~\GeV^2$ & $|t|=0.5~\GeV^2$ & $|t|=1.0~\GeV^2$ \\
\midrule
$\widehat N_0(t)$ & $0.496$ & $0.233$ & $0.098$ \\
$\widehat N_2^{A}(t)$ & $0.770$ & $0.545$ & $0.334$ \\
$\widehat N_2^{A,J}(t)$ & $0.758$ & $0.519$ & $0.295$ \\
$\widehat N_2^{A,J,D}(t)$ & $0.718$ & $0.482$ & $0.269$ \\
\bottomrule
\end{tabular}
\end{table}

\begin{figure}[H]
\centering
\includegraphics[width=0.66\textwidth]{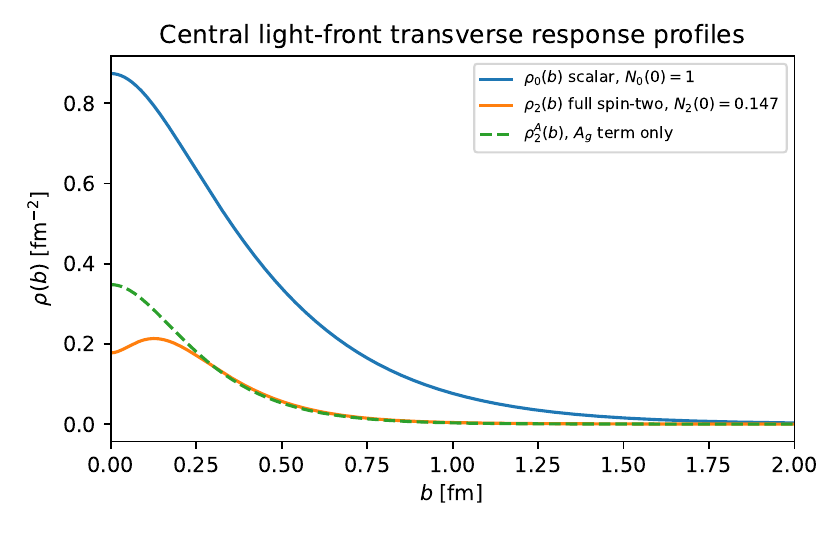}
\caption{Central light-front transverse response profiles obtained from Eq.~\eqref{eq:profile_defs}.  The dashed curve shows the profile obtained if the spin-two numerator is approximated by the $A_g(t)$ term alone.}
\label{fig:profiles}
\end{figure}

The corresponding light-front transverse profiles are shown in Fig.~\ref{fig:profiles}.  The plotted normalizations are $N_0(0)=1$ and $N_2(0)=0.147$.  The spin-two profile is broader in this benchmark because the scalar trace shape decreases more rapidly with $|t|$.  The full spin-two profile is not simply a rescaled $A_g$ profile; the $J_g$ and $D_g$ terms suppress the near-origin response and shift weight to finite transverse distance.  The figure should be read as a response-profile comparison, not as a probability-density decomposition of the proton.

\FloatBarrier

\section{Cumulative ratio and robustness}
\label{sec:ratio}

\begin{figure}[H]
\centering
\includegraphics[width=0.66\textwidth]{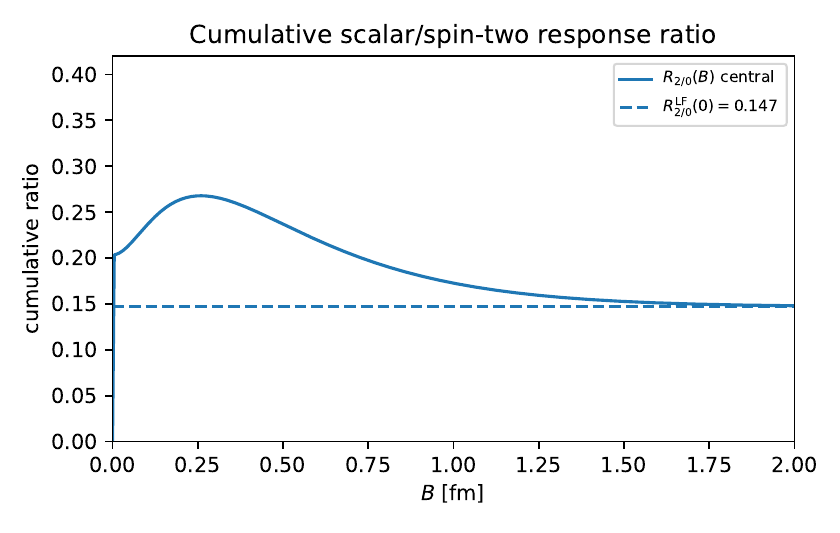}
\caption{Central cumulative ratio $R_{2/0}(B)$.  The horizontal line is the forward light-front spin-two/scalar amplitude ratio.}
\label{fig:cumulative}
\end{figure}

The cumulative ratio of Eq.~\eqref{eq:cumulative} converts the profile comparison into a single scale-dependent diagnostic.  Figure~\ref{fig:cumulative} shows the central result.  At finite $B$ the ratio can exceed its asymptotic value because the spin-two response is broader and develops more weight at intermediate transverse distances than the scalar profile in the central benchmark.  The ratio then approaches the externally fixed forward value $R_{2/0}^{\rm LF}(0)=0.147$.

Selected central values of the cumulative ratio are given in Table~\ref{tab:cumulative_numbers}.

\begin{table}[H]
\centering
\caption{Central cumulative ratio at selected transverse integration radii.}
\label{tab:cumulative_numbers}
\begin{tabular}{lcccccc}
\toprule
$B$ [fm] & $0.3$ & $0.5$ & $0.7$ & $1.0$ & $2.0$ & $3.0$ \\
\midrule
$R_{2/0}(B)$ & $0.266$ & $0.237$ & $0.204$ & $0.173$ & $0.148$ & $0.147$ \\
\bottomrule
\end{tabular}
\end{table}

\begin{figure}[H]
\centering
\includegraphics[width=0.66\textwidth]{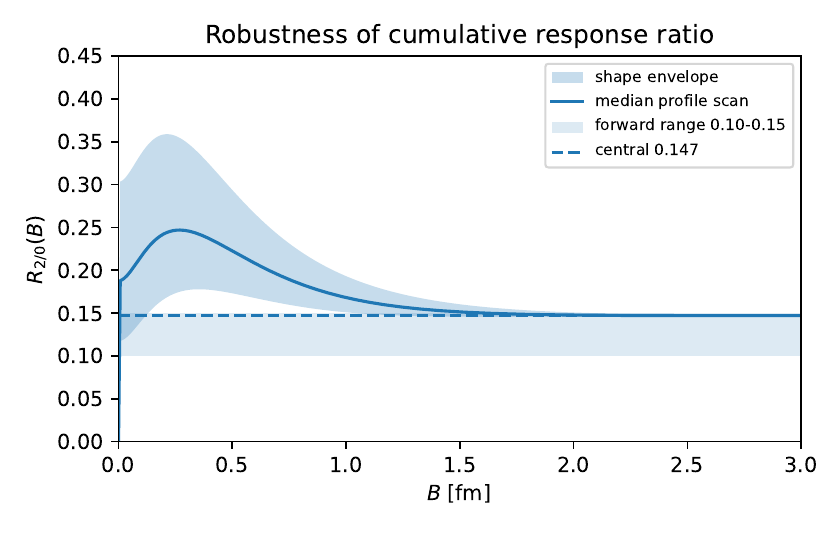}
\caption{Robustness scan of the cumulative ratio.  The envelope varies the scalar profile and the $D$-term normalization at fixed central forward strength.  The pale horizontal band shows the independent forward-scale range $R_{2/0}^{\rm LF}(0)\simeq0.10$--$0.15$.}
\label{fig:robustness}
\end{figure}

The finite-$B$ approach to the asymptotic ratio is model dependent.  Figure~\ref{fig:robustness} scans the scalar slope parameter $r_s=0.75,0.85,0.97,1.10~\fm$ and rescales the gluon $D$ term by $\eta_D=0,0.5,1,1.5$.  The shaded envelope shows shape uncertainty at the central forward ratio.  The additional horizontal band indicates the forward normalization range $R_{2/0}^{\rm LF}(0)\simeq0.10$--$0.15$ generated by the matching-scale variation quoted in Ref.~\cite{Syamtomov2026EMT}.  Thus the asymptote is controlled by the forward chromoelectric matching, whereas the saturation rate is controlled by the relative scalar and gravitational transverse profiles.

\FloatBarrier

\section{Interpretation and conclusions}
\label{sec:conclusions}

The light-front profile construction gives a clean spatial language for the chromoelectric EMT response.  Its main use is not to replace a full near-threshold production analysis, but to keep separate the ingredients that enter such an analysis: scalar normalization, spin-two normalization, off-forward scalar shape, off-forward gravitational shape, and possible process-dependent phases in the physical amplitude.

The construction has three practical features.  First, the scalar response is the anomaly and sigma amplitude of the matched chromoelectric operator, not a pure gluonic trace.

Second, the spin-two profile is built from the contraction
\begin{equation}
3B_i-D_i=6J_i-3A_i-D_i,
\end{equation}
not from $A_i(t)$ alone.  Third, the scheme-dependent $\bar C_i(t)$ term is not assigned its own chromoelectric response density because it cancels from the traceless spin-two contraction.

The numerical benchmark shows that a forward spin-two correction of order $10$--$15\%$ can correspond to a more visible finite-$B$ effect if the scalar profile is more localized than the spin-two profile.  This is a shape effect, not a violation of scalar dominance.  The cumulative ratio always returns to the forward light-front value at large $B$, while its approach to that limit encodes the relative transverse scales.

Near-threshold charmonium observables therefore probe a chromoelectric projection of scalar and spin-two EMT responses.  Differential measurements can constrain the relative transverse slopes and the spin-two combination $6J_i-3A_i-D_i$, but they should not be interpreted as measuring a unique model-independent static mass radius of the proton.

\end{document}